\documentclass[12pt]{article}

\newtheorem{proposition}{Proposition}

\makeatletter
\newcommand*{\indep}{%
\mathbin{%
\mathpalette{\@indep}{}%
}%
}
\newcommand*{\nindep}{%
\mathbin{% % The final symbol is a binary math operator
\mathpalette{\@indep}{\not}% \mathpalette helps for the adaptation
}%
}
\newcommand*{\@indep}[2]{%
\sbox0{$#1\perp\m@th$}% box 0 contains \perp symbol
\sbox2{$#1=$}% box 2 for the height of =
\sbox4{$#1\vcenter{}$}% box 4 for the height of the math axis
\rlap{\copy0}% first \perp
\dimen@=\dimexpr\ht2-\ht4-.2pt\relax
\kern\dimen@
{#2}%
\kern\dimen@
\copy0 % second \perp
}

\usepackage[figuresright]{rotating}
\usepackage{bm}
\usepackage{amsmath,dsfont}
\usepackage{graphicx,booktabs}
\usepackage{natbib}
\usepackage[ margin=1.11in]{geometry}

\makeatletter
\newcommand*\bigcdot{\mathpalette\bigcdot@{.5}}
\newcommand*\bigcdot@[2]{\mathbin{\vcenter{\hbox{\scalebox{#2}{$\m@th#1\bullet$}}}}}
\makeatother
\title{Doubly Robust Estimation in Observational Studies with Partial Interference}

\author
{
Lan Liu, Michael G. Hudgens, Bradley Saul, John D. Clemens, \\Mohammad Ali, Michael E. Emch}
\begin{document}

%  This will produce the submission and review information that appears
%  right after the reference section.  Of course, it will be unknown when
%  you submit your paper, so you can either leave this out or put in 
%  sample dates (these will have no effect on the fate of your paper in the
%  review process!)

\maketitle

%  put the summary for your paper here
%\newpage
\begin{abstract}
Interference occurs when the treatment (or exposure) of one individual affects the outcomes of others. In some settings it may be reasonable to assume individuals can be partitioned into clusters such that there is no interference between individuals in different clusters, i.e., there is partial interference. 
In observational studies with partial interference, inverse probability weighted (IPW) estimators have been proposed of different possible treatment effects. However, the validity of IPW estimators depends on the propensity score being known or correctly modeled. Alternatively, one can estimate the treatment effect using an outcome regression model. In this paper, we propose
doubly robust (DR) estimators which utilize both models and are consistent and asymptotically normal if either model, but not necessarily both, is correctly specified. Empirical results are presented to demonstrate the DR property of the proposed estimators, as well as the efficiency gain of DR over IPW estimators when both models are correctly specified. The different estimators are illustrated using data from a  study examining the effects of cholera vaccination in Bangladesh.
\end{abstract}

%\begin{keywords}
%Causal inference; Doubly robust estimator; Interference; Observational studies.\end{keywords}

\section{Introduction}
%\vspace{-0.2cm}
Typically in causal inference it is assumed an individual's potential outcomes do not depend on the treatment (or exposure) of other individuals, i.e., there is no interference \citep{cox1958planning}. However, this assumption may not hold in various settings. For example, in a vaccine trial, the infection status of one individual may depend on whether other individuals are vaccinated. Interference may occurs in other areas, such as econometrics \citep{sobel2006randomized, manski2013identification}, education \citep{hong2006evaluating,Basse2018analyzing}, and political science \citep{sinclair2012detecting,bowers2013reasoning}. 

Recently, inference methods have been proposed for settings where individuals can be partitioned into clusters and possible interference exists only among individuals in the same cluster. This is sometimes called partial interference \citep{sobel2006randomized} and can be viewed as a special case of the constant treatment response assumption \citep{manski2013identification}. \cite{hudgens2008toward} proposed estimators of direct, indirect
(or spillover), total, and overall causal effects of a treatment for two-stage randomized experiments, and \cite{liu2013asymptotic} derived the asymptotic distributions of these estimators. \cite{tchetgen2012causal} proposed inverse probability weighted (IPW) estimators of these causal effects for observational studies. However, the validity of these IPW estimators only holds when the propensity score is known or correctly modeled. Moreover, IPW estimators are known to have large variances and be unstable, especially when some propensity scores are close to 0 or 1, which may be common when there is partial interference. %Thus, new methods need to be developed to improve the efficiency and stabilize IPW estimators.

In the absence of interference, doubly robust (DR) estimators are known to have certain advantages over IPW estimators. DR estimators are constructed by utilizing two models: a model for the dependence of treatment on covariates (i.e., propensity score model), and a model for the dependence of the outcome on covariates and treatment. DR estimators are consistent when either, but not necessarily both, of the two models is correct. In practice, neither the model for the propensity score nor the outcome model is known. Thus, a DR estimator provides two chances to consistently estimate the parameter of interest. However, existing DR estimators assume no interference and hence are not applicable in settings such as infectious diseases where interference may be present.   

In this paper, several DR estimators are proposed for use in observational studies where there may be partial interference. The outline of the remainder of the paper is as
follows. In Section \ref{sec: DR_notation}, notation, assumptions, and the causal effects of interest are introduced. The IPW and regression estimators are defined in Section \ref{sec: IPW_and_reg_estimator} and various DR estimators are proposed in Section \ref{sec: DE_estimator}. Results
from a simulation study are presented in Section \ref{sec: DR_simulation}. The proposed DR estimators are used to analyze data from a cholera vaccine study in Section \ref{sec: DR_dia}. Finally, Section \ref{sec: DR_discussion} concludes with a discussion. 

\section{Notation, Assumptions and Estimands}\label{sec: DR_notation}

\vspace{-0.2cm}
Consider an observational study where data is observed for individuals who can be partitioned into groups (e.g., students in different schools). Suppose there are $k$ groups of individuals in the study with $N_i$ individuals in group $i$. For individual $j$ in group $i$ we observe $(X_{ij},A_{ij},Y_{ij})$ for $j=1,\ldots,N_i$, $i=1,\ldots,k$, where $X_{ij}$ denotes a vector of pre-treatment covariates, $A_{ij}$ denotes a treatment indicator ($A_{ij}=1$ if individual receives treatment and $A_{ij}=0$ otherwise), and $Y_{ij}$ is a univariate outcome of interest, which can be continuous or categorical. Let $X_i=(X_{i1},\ldots,X_{iN_i})$, $A_i=(A_{i1},\ldots,A_{iN_i})$ and $Y_i=(Y_{i1},\ldots,Y_{iN_i})$. Assume the $k$ groups are a random sample from an infinite super-population of groups such that $O_i=(X_i,A_i,Y_i)$ are independent and identically distributed for $i=1,\ldots,k$. Define $A_{i(-j)}=A_i\backslash A_{ij}$, i.e., the vector of treatment indicators for all individuals in group $i$ except individual $j$. Let $a_{ij}$, $a_{i(-j)}$ and $a_i$ denote  possible realizations of $A_{ij}$, $A_{i(-j)}$ and $A_i$.  Define $f(a_i|x_i)=\Pr(A_i=a_i|X_i=x_i)$ to be the probability of treatment vector $a_i$ given covariates $x_i$ and similarly define $f(a_{ij}|x_i)=\Pr(A_{ij}=a_{ij}|X_i=x_i)$. Assume $f(a_i|x_i)>0$ for all $x_i$ in the support of $X_i$; this is sometimes referred to as the positivity assumption.

Assume there is no interference between individuals in different groups, i.e., partial interference. This assumption may be reasonable in settings where groups are sufficiently separated geographically or in time. Note no assumption is made about the nature of interference within groups. Indeed one of the primary inferential goals is to assess to what extent there is interference within groups. Assuming partial interference, the potential outcome of one individual may be expressed as a function of their own treatment as well as the treatment of others in the same group.  Therefore, the potential outcome for individual $j$ in group $i$ is denoted $Y_{ij}(a_i)=Y_{ij}(a_{ij},a_{i(-j)})$ for treatment vector $a_i$. Additionally, we make the causal consistency assumption that the observed outcome $Y_{ij}$ is the same as the potential outcome $Y_{ij}(a_i)$ if treatment $A_i=a_i$, i.e., $Y_{ij}=\sum_{a_i}1(A_{i}=a_i)Y_{ij}(a_i)$.
 Assume $Y_{i}(a_{i})\indep A_i|X_i$, where $\indep$ denotes independence; this assumption is sometimes referred to as conditional exchangeability or ignorability.  
%$f(A_i|X_i)=f(A_i|X_i,Y_{ij}(\cdot))$

Causal effects of treatment are defined by average outcomes under different counterfactual scenarios corresponding to different distributions of treatment in the population. Following \cite{tchetgen2012causal}, consider the treatment allocation strategy (or policy) where individuals receive treatment independently with probability $\alpha$. Under an $\alpha$ allocation strategy, the probability of treatment $A_i=a_i$ for group $i$ is $\pi(a_i;\alpha)=\Pr_{\alpha}(A_i=a_i)=\prod_{j} \alpha^{a_{ij}}(1-\alpha)^{1-a_{ij}}$. The $\alpha$ subscript of $\Pr_{\alpha}$ indicates probability in the counterfactual scenario corresponding to policy $\alpha$. Similarly, let $\pi(a_{i(-j)};\alpha)=\Pr_{\alpha}(A_{i(-j)}=a_{i(-j)})=\prod_{k\neq j} \alpha^{a_{ik}}(1-\alpha)^{1-a_{ik}}$ denote the probability of treatment $A_{i(-j)}=a_{i(-j)}$ for all individuals in group $i$ other than individual $j$. Define the average potential outcome under policy $\alpha$ when an individual receives treatment $a$ by $ \mu_{a\alpha}= \sum_{a_{i(-j)}}E \{Y_{ij}(a,a_{i(-j)})\}\pi(a_{i(-j)};\alpha)$,  where $E\{\cdot\}$ denotes the expected value in the super-population of groups. Similarly, define the average potential outcome under policy $\alpha$ to be $\mu_{\alpha}= \sum_{a_i}E\{Y_{ij}(a_i)\}\pi(a_i;\alpha)$. Following \citet{halloran1995causal} and \citet{hudgens2008toward}, define the direct effect of treatment under policy $\alpha$ to be $\overline{DE}(\alpha)=\mu_{1\alpha}-\mu_{0\alpha}$. For policies $\alpha_0$ and $\alpha_1$, define the indirect effect 
$\overline{IE}(\alpha_1,\alpha_0)= \mu_{0\alpha_1}-\mu_{0\alpha_0}$, the total effect
$\overline{TE}(\alpha_1,\alpha_0)=  \mu_{1\alpha_1}-\mu_{0\alpha_0}$, and the overall effect
$\overline{OE}(\alpha_1,\alpha_0)=  \mu_{\alpha_1}-\mu_{\alpha_0}$.
In words, the direct effect is the difference between the average potential outcomes when group $i$ receives policy  $\alpha$ and an individual in that group receives treatment compared to when an individual in that group receives control. The indirect (or spillover) effect compares the average potential outcome when an individual receives control under different policies $\alpha_1$ and $\alpha_0$. The total effect equals
the sum of the direct and indirect effects,
and the overall effect provides a single summary measure of the effect of policies $\alpha_1$ versus $\alpha_0$.
See \citet{tchetgen2012causal} for further discussion about these estimands. %These estimands characterize the various effects of treatment in the presence of interference.
%\vspace{-0.2cm}
\section{IPW and Regression Estimators}\label{sec: IPW_and_reg_estimator}

Inverse probability weighting is a common approach to adjusting for confounding in observational studies. Heuristically, inverse probability weighting creates a pseudo-population in which there is no confounding such that the average outcome in the pseudo-population approximates the average outcome that would have been observed if treatment has been randomly assigned. 
%Under the no unmeasured confounders assumption, 
%%
%%\vspace{-6mm}\begin{eqnarray*}
%%f\{A_i|f(A_i|X_i),Y_{ij}(a_i)\}&=&E[f\{A_i|X_i,Y_{ij}(a_i)\}|f(A_i|X_i),Y_{ij}(a_i)]\\
%%&=&E\{f(A_i|X_i)|f(A_i|X_i),Y_{ij}(a_i)\}\\
%%&=&f(A_i|X_i)=f(A_i|f(A_i|X_i)).
%%\end{eqnarray*}
%%\noindent Thus, 
%it is straightforward to show that $Y_{ij}({a_i})\indep A_i|f(A_i|X_i)$, i.e., we only need to control for the propensity score $f(A_i|X_i)$ to control for the confounding.
 \citet{tchetgen2012causal} proposed IPW estimators for $\mu_{a\alpha}$ and $\mu_{\alpha}$ defined by $\widehat{Y}^{ipw}(a;\alpha)= \sum_{i=1}^k\widehat{Y}_i^{ipw}(a;\alpha)/k$ and $\widehat{Y}^{ipw}(\alpha)= \sum_{i=1}^k\widehat{Y}_i^{ipw}(\alpha)/k$, where 
\[\widehat{Y}_i^{ipw}(a;\alpha)= N_i^{-1} \sum_{j=1}^{N_i}1(A_{ij}=a)Y_{ij}(A_{i})\pi(A_{i(-j)};\alpha)/{f}(A_i|X_i;\hat\gamma),\]  \[\widehat{Y}_i^{ipw}(\alpha)= N_i^{-1} \sum_{j=1}^{N_i}Y_{ij}(A_{i})\pi(A_i;\alpha_0)/{f}(A_i|X_i;\hat\gamma),\] and ${f}(A_i|X_i;\gamma)$ denotes a propensity score model with finite-dimentional vector of parameters $\gamma$ and $\hat \gamma$ is an estimator of $\gamma$. IPW estimators of the direct, indirect, total and overall effect are then defined as  $\widehat{\textsc{de}}^{ipw}(\alpha)=\widehat{Y}^{ipw}(1,\alpha)-\widehat{Y}^{ipw}(0,\alpha)$, $\widehat{\textsc{ie}}^{ipw}(\alpha_1,\alpha_0)=\widehat{Y}^{ipw}(0,\alpha_1)-\widehat{Y}^{ipw}(0,\alpha_0)$, $\widehat{\textsc{te}}^{ipw}(\alpha_1,\alpha_0)=\widehat{Y}^{ipw}(1,\alpha_1)-\widehat{Y}^{ipw}(0,\alpha_0)$ and $\widehat{\textsc{oe}}^{ipw}(\alpha_1,\alpha_0)=\widehat{Y}^{ipw}(\alpha_1)-\widehat{Y}^{ipw}(\alpha_0)$, respectively. Assuming a correctly specified mixed effects logistic regression model for the propensity score $f(A_{ij}|X_i;\gamma)$ and $\hat \gamma$ equal to the maximum likelihood estimator of $\gamma$, \citet{PerezHeydrich2014interference} proved the IPW estimators are consistent and asymptotically normal by showing the estimators solve a vector of unbiased estimating equations.%The consistency and asymptotic normality of the IPW causal effects estimators can be found in  \citet{PerezHeydrich2014interference}. 

Alternatively, one can adjust for confounding by controlling for observed covariates in an outcome regression model $m_{ij}(a_i,X_i;\beta)=E\{Y_{ij}(a_i)|X_i;\beta\}$, where $\beta$ is a finite-dimensional vector of model parameters in the outcome regression model. By the exchangeability assumption, \[ m_{ij}(a_i,X_i;\beta)=E\{Y_{ij}(a_i)|A_i=a_i,X_i;\beta\}=E(Y_{ij}|A_i=a_i,X_i;\beta);\] thus, model parameters are identifiable based on the observable random variables $O_i$. For example, a regression model for $Y_{ij}(A_i)$ is $m_{ij}(A_i,X_i;\beta)=\beta_1+\beta_{A_{ij}}A_{ij}+\beta_{A_{i(-j)}}^TA_{i(-j)}+\beta_{X_i}^{T}X_i$. Define  $\widehat{Y}_i^{reg}(a;\alpha)=\sum_{j=1}^{N_i}\sum_{a_{i(-j)}}m_{ij}(a,a_{i(-j)},X_i;\hat\beta)\pi(a_{i(-j)};\alpha)/N_i$ and $\widehat{Y}^{reg}_i(\alpha)=\sum_{j=1}^{N_i}\sum_{a_i}m_{ij}(a_i,X_i;\hat\beta)\pi(a_i;\alpha)/N_i$, where $\hat{\beta}$ is the least squares estimator for $\beta$. Define the regression estimators of $\mu_{a\alpha}$ and $\mu_{\alpha}$ to be $\widehat{Y}^{reg}(a;\alpha)=\sum_{i=1}^k\widehat{Y}_i^{reg}(a;\alpha)/k$ and $\widehat{Y}^{reg}(\alpha)=\sum_{i=1}^k\widehat{Y}_i^{reg}(\alpha)/k$, with the corresponding regression causal effect estimators defined analogously to the IPW causal effect estimators defined above. Similar to the IPW estimators, it is straightforward to show that if the outcome regression model is correctly specified, then $\widehat{Y}^{reg}(a;\alpha)$ and $\widehat{Y}^{reg}(\alpha)$ are consistent and asymptotically normal estimators of $\mu_{a\alpha}$ and $\mu_{\alpha}$ using standard estimating equation theory.

%The regression causal effect estimators could be defined similarly as the IPW estimators. Similar to the IPW estimators, it is straightforward to show under correct model specification that the regression causal effect estimators  are consistent and asymptotically normal using standard estimating equation theory. %The asymptotic normality and variances of the regression causal effect estimators could be derived using the M-estimator theory. 

Thus, the various causal effects defined above can be consistently estimated by the IPW estimator if the propensity score model is correctly specified. These effects  can also be consistently estimated by the outcome regression estimator if the regression model is correctly specified. In the next section, several DR estimators are proposed which utilize both the propensity score and regression models, and are consistent if either model (but not necessarily both) is correctly specified.

\vspace{-0.3cm}
\section{Doubly Robust Estimators}\label{sec: DE_estimator}

\subsection{Regression estimation with residual bias correction}
Define $ \widehat{Y}^{DR\bigcdot BC}(a,\alpha)= \sum_{i=1}^k\widehat{Y}_i^{DR\bigcdot BC}(a,\alpha)/k$ and $ \widehat{Y}^{DR\bigcdot BC}(\alpha)= \sum_{i=1}^k\widehat{Y}_i^{DR\bigcdot BC}(\alpha)/k$ to be the residual bias correction DR estimators for $\mu_{a\alpha}$ and $\mu_{\alpha}$, where

\vspace{-0.8cm}
\begin{eqnarray*}\label{eq: DR_estimator}
  \widehat{Y}_i^{DR\bigcdot BC}(a,\alpha)&=&N_i^{-1}\sum_{j=1}^{N_i}\Biggl\{ \sum_{a_{i(-j)}}m_{ij}(a,a_{i(-j)},X_i;\hat\beta)\pi(a_{i(-j)};\alpha)\\&&+\frac{1(A_{ij}=a)}{{f}(A_i|X_i;\hat\gamma)}
  \{Y_{ij}(A_i)-m_{ij}(A_i,X_i;\hat\beta)\}\pi(A_{i(-j)};\alpha) 
 \Biggr\},
 \end{eqnarray*}

 \vspace{-0.6cm}
 \begin{equation*}
  \widehat{Y}_i^{DR\bigcdot BC}(\alpha)=N_i^{-1}\sum_{j=1}^{N_i}\left\{
 \sum_{a_i}m_{ij}(a_i,X_i;\hat\beta)\pi(a_i;\alpha)+ \frac{\{Y_{ij}(A_i)-m_{ij}(A_i,X_i;\hat\beta)\}}{{f}(A_i|X_i;\hat\gamma)}\pi(A_i;\alpha)\right\}.
\end{equation*}

\noindent The bias correction DR estimators are motivated by the DR estimators proposed by \cite{scharfstein1999adjusting} for the setting where there is no interference. The bias correction DR estimators are composed of two parts. The first part is the regression estimator and the second part entails inverse weighted residuals of the regression estimator. Informally, the DR property of these estimators follows by noting: (i) when the regression estimator is correctly specified, the first part is consistent for the parameter of interest and the second part converges to 0; (ii) when the regression estimator is misspecified but the propensity score model is correctly specified, the first part is biased but the second part consistently estimates the bias of the first term such that the summation is still consistent for the target parameter.%thus the summation is still consistent. %A more rigorous proof of the DR property of the bias correction estimator is given in the Appendix.%The DR property of the bias correction estimators are given in the Proposition \ref{thm: DR_CAN}, the proof of which is relegated to the Appendix. 

The bias correction DR causal effect estimators are defined similarly to the IPW causal effect estimators in Section \ref{sec: IPW_and_reg_estimator}. For example, the bias correction DR direct effect estimator is $\widehat{\textsc{de}}^{DR\bigcdot BC}(\alpha)=\widehat{Y}^{DR\bigcdot BC}(1,\alpha)-\widehat{Y}^{DR\bigcdot BC}(0,\alpha)$. To derive the asymptotic distribution of the bias correction direct effect estimator, let $G^{DR\bigcdot BC}_{a\alpha}(O_i;\mu,\beta,\gamma) =\widehat{Y}_i^{DR\bigcdot BC}(a,\alpha)- \mu$ and let $G^{DR\bigcdot BC}_{\beta}(O_i;\beta)$ and $G^{DR\bigcdot BC}_{\gamma}(O_i;\gamma)$ denote the estimating functions corresponding to $\hat \beta$ and $\hat \gamma$, such that $\hat\theta^{DR\bigcdot BC}=\{\hat Y^{DR\bigcdot BC}(0,\alpha),\hat Y^{DR\bigcdot BC}(1,\alpha),\hat\beta,\hat\gamma\}$ is the solution to the vector equation $\sum_{v=1}^k G^{D,DR\bigcdot BC}_{\alpha}(O_i;\theta)=0$ where $\theta=(\mu_{0\alpha},\mu_{1\alpha},\beta,\gamma)$ and $G^{D,DR\bigcdot BC}_{\alpha}(O;\theta)=\{G^{DR\bigcdot BC}_{0\alpha}(O;\mu_{0\alpha},\beta,\gamma),$ $G^{DR\bigcdot BC}_{1\alpha}(O;\mu_{1\alpha},\beta,\gamma),$ $G^{DR\bigcdot BC}_{\beta}(O;\beta),$ $ G^{DR\bigcdot BC}_{\gamma}(O;\gamma)\}^T$.  The following proposition shows the DR property and the asymptotic normality of the bias correction DR estimator for the direct effect; the proof is in the Appendix. The DR property and asymptotic normality for the other bias correction DR causal effect estimators can be derived similarly.

%, that is, $\beta^{\ast}=\beta_0$ or $\gamma^{\ast}=\gamma_0$

\begin{proposition}\label{thm: DR_CAN}
If either $f(A_i|X_i;\gamma)$ or $m_{ij}(A_i,X_i;\beta)$ is correctly specified, then \\\emph{$k^{1/2}\{\widehat{\textsc{de}}^{DR\bigcdot BC}(\alpha)-\overline{\textsc{de}}(\alpha)\}$} converges in distribution to $N(0,\Sigma_0^{D})$ as $k \to \infty$ where
 $
\Sigma^{D} = \tau U^{-1} V U^{-T}\tau^T,$
$U=-E\{\partial G_{\alpha}^{D,DR\cdot BC}(O_i;\theta)/\partial \theta\}$, $V=E\{G_{\alpha}^{D,DR\bigcdot BC}(O_i;\theta)^{\otimes 2}\}$ and $\tau=(1,-1, 0,\ldots,0)$.
\end{proposition}

A consistent estimator of the asymptotic variance of $\widehat{\textsc{de}}^{DR\bigcdot BC}(\alpha)$ can be constructed by replacing expectations in $U$ and $V$ with their empirical counterparts. Consistent variance estimators of other bias correction DR causal effect estimators can be constructed similarly.

In practice,  the summation terms of the form $ \sum_{a_i}m_{ij}(a_i,X_i;\hat\beta)\pi(a_i;\alpha)$ in the bias correction DR estimators may be time consuming to calculate since the summation is over all possible value of $a_i$. However, a Monte Carlo  approximation can be employed by: (i) independently sampling $\tilde A_{ij}$ from a Bernoulli distribution with mean $\alpha$ for $j=1,\ldots,N_i$; (ii) calculating $m_{ij}((\tilde A_{i1},\ldots,\tilde A_{iN_i}),X_i;\hat \beta)$; (iii) repeating steps (i) and (ii) $MC$ times; and (iv) averaging the $MC$ values of $m_{ij}((\tilde A_{i1},\ldots,\tilde A_{iN_i}),X_i;\hat \beta)$. This will provide an unbiased estimate of $\sum_{a_i}m_{ij}(a_i,X_i;\hat\beta)\pi(a_i;\alpha)$, with larger values of $MC$ resulting in smaller variability of the approximation.

%However, a Monte Carlo approximation can be made by sampling the treatment $\tilde A_{ij}$ independently from a Bernounlli distribution with mean $\alpha$ and calculating $m_{ij}(a_i,X_i;\hat\beta)$ by averaging the $m_{ij}(\tilde a_i,X_i;\hat\beta)$ over the sampled treatment where $\tilde a_{i}=a_i$. This provides an unbiased estimate for $\sum_{a_i}m_{ij}(a_i,X_i;\hat\beta)\pi(a_i;\alpha)$. More Bernoulli samples will result in smaller variability of such approximation. 

\subsection{Regression estimation with inverse-propensity weighted coefficients}\label{subsec: DR_WLS}

In this section we consider a second DR estimator which can be viewed as a generalization of the weighted least squares estimator in \citet{kang2007demystifying} to the partial interference setting. Let $L_{ij}=(1,A_{i(-j)},X_i)$ denote the row vector of all regressors including the intercept in the outcome regression model when $A_{ij}=a$, which for simplicity, we write as $m_{ij}(a,A_{i(-j)},X_i;\beta)=m_{ij}(a,L_i;\beta)$, where  $L_i=(L_{i1}^T,\ldots,L_{iN_i}^T)^T$. Let $ m_i=(m_{i1},\ldots,m_{iN_i})$ and note the parameter $\beta$ is the solution to the equation $\int G^{reg}(O;\beta)dF(o) =0$ where $$G^{reg}(O_i;\beta)=L_i^T\Lambda_i(A_i,X_i,\omega_{i})\{Y_i- m_i(a,L_i;\beta)\}^T,$$ 
  $\Lambda_i(A_i,X_i,\omega_{i})=\text{diag}\biggl\{1(A_{i1}=a)\omega_{i1}(L_{i}),\ldots,1(A_{iN_i}=a)\omega_{iN_i}(L_{i})\biggr\}$ for any user specified vector-valued  function $\omega_{i}=(\omega_{i1},\ldots,\omega_{iN_i})$ and  in general diag$(x_1,\ldots,x_n)$ denotes an $n \times n$ diagnoal matrix with entries $x_1,\ldots,x_n$ along the diagonal. The choice $\omega_{ij}=1$ corresponds to the normal equations of the standard least squares estimator. To achieve the DR property, we use 
  $$\omega_i^{WLS}(L_i;\alpha,\gamma)=\biggl\{\frac{\pi(A_{i(-1)};\alpha)}{{f}(a,A_{i(-1)}|X_i;\gamma)}
  ,\ldots,\frac{\pi(A_{i(-N_i)};\alpha)}{{f}(a,A_{i(-N_i)}|X_i;\gamma)}
  \biggr\},$$

\noindent and let $$G^{reg\bigcdot WLS}_{\alpha}(O_i;\beta,\gamma)=L_i^T\Lambda_i(A_i,X_i,\omega_{i}^{WLS};\alpha,\gamma)\{Y_i- m_i(a,L_i;\beta)\}^T.$$ As shown below, this construction yields another DR estimator. Define the weighted coefficients DR estimator by

\vspace{-0.8cm}
\begin{eqnarray*}\label{eq: DR_WLS_estimator}
  \widehat{Y}_i^{DR\bigcdot WLS}(a,\alpha)&=&N_i^{-1}\sum_{j=1}^{N_i}\Biggl\{\sum_{a_{i(-j)}}m_{ij}(a,a_{i(-j)},X_i;\hat\beta^{WLS})\pi(a_{i(-j)};\alpha)
 \Biggr\},
 \end{eqnarray*}

\noindent where $\hat\beta^{WLS}=\hat\beta^{WLS}(\alpha)$ is obtained by solving 

\begin{equation}\label{eq: ee_DR_WLS}
\sum_{i=1}^kG^{reg\bigcdot WLS}_{\alpha}(A_i,X_i;\beta,\hat\gamma)=0.
\end{equation}

\noindent Define $\widehat{Y}_i^{DR\bigcdot WLS}(\alpha)$ similarly. The population level estimators and hence the causal effect estimators can be obtained by averaging the group level estimators as before. 

%  \widehat{Y}_i^{DR\bigcdot BC}(a,\alpha)&=&N_i^{-1}\sum_{j=1}^{N_i}\Biggl\{ \sum_{a_{i(-j)}}m_{ij}(a,a_{i(-j)},X_i;\hat\beta)\pi(a_{i(-j)};\alpha)\\&&+\frac{1(A_{ij}=a)}{{f}(A_i|X_i;\hat\gamma)}
%  \{Y_{ij}(A_i)-m_{ij}(A_i,X_i;\hat\beta)\}\pi(A_{i(-j)};\alpha) 
% \Biggr\},

To show the DR property of the weighted coefficients DR estimators, notice \eqref{eq: ee_DR_WLS} implies

\begin{equation}\label{eqn: key_ee_WLS}
N_i^{-1}\sum_{j=1}^{N_i}\Biggl[ \frac{1(A_{ij}=a)}{{f}(A_i|X_i;\hat\gamma)}
  \{Y_{ij}(A_i)-m_{ij}(A_i,X_i;\hat\beta^{WLS})\}\pi(A_{i(-j)};\alpha) 
 \Biggr]=0.
\end{equation}

\noindent Thus, the weighted coefficients DR estimator can be written as
\begin{eqnarray*}
  \widehat{Y}_i^{DR\bigcdot WLS}(a,\alpha)&=&N_i^{-1}\sum_{j=1}^{N_i}\Biggl\{\sum_{a_{i(-j)}}m_{ij}(a,a_{i(-j)},X_i;\hat\beta^{WLS})\pi(a_{i(-j)};\alpha)
 \Biggr\}\\
 &=&N_i^{-1}\sum_{j=1}^{N_i}\Biggl\{\sum_{a_{i(-j)}}m_{ij}(a,a_{i(-j)},X_i;\hat\beta^{WLS})\pi(a_{i(-j)};\alpha)\\
 &&\quad+\frac{1(A_{ij}=a)}{{f}(A_i|X_i;\hat\gamma)}
  \{Y_{ij}(A_i)-m_{ij}(A_i,X_i;\hat\beta^{WLS})\}\pi(A_{i(-j)};\alpha)
 \Biggr\},
 \end{eqnarray*}

\noindent which has the same form as the bias correction DR estimator and the DR property can be shown in a similar fashion. In particular, let $\hat \theta^{DR\bigcdot WLS}=\{\hat Y^{DR\bigcdot WLS}(0,\alpha),\hat Y^{DR\bigcdot WLS}(1,\alpha),\hat\beta,\hat\gamma\}$, which is the solution to the estimating equation
$\sum_{i=1}^k G^{D,DR\bigcdot WLS}_{\alpha}(O_i;\theta)=0$, where \\$G^{ D,DR\bigcdot WLS}_{\alpha}(O;\theta) = \{G^{DR\bigcdot WLS}_{0\alpha}(O;\mu_{0\alpha},\beta,\gamma),G^{DR\bigcdot WLS}_{1\alpha}(O;\mu_{1\alpha},\beta,\gamma),G^{reg\bigcdot WLS}_{\alpha}(O;\beta,\gamma),G^{ipw}(A,X;\gamma)\}^T$ and $G^{DR\bigcdot WLS}_{a\alpha}(O;\mu_{a\alpha},\beta,\gamma)=\widehat Y_i^{DR\bigcdot WLS}(a,\alpha)-\mu$.
The DR property and asymptotic normality of the weighted direct effect estimator are formally stated in the following proposition.

\begin{proposition}\label{prop: DR_WLS}
If either $f(A_i|X_i;\gamma)$ or $m_{ij}(A_i,X_i;\beta)$ is correctly specified, then \\\emph{$k^{1/2}\{\widehat{\textsc{de}}^{DR\bigcdot WLS}(\alpha)-\overline{\textsc{de}}(\alpha)\}$} converges in distribution to $N(0,\Sigma_0^{D})$ as $k \to \infty$ where 
$
\Sigma^{D} = \tau U^{-1} V U^{-T}\tau^T,
$
$U=-E\{\partial G_{\alpha}^{D,DR\bigcdot WLS}(O_i;\theta)/\partial \theta\}$, $V=E\{G_{\alpha}^{D,DR\bigcdot WLS}(O_i;\theta)^{\otimes 2}\}$, $\tau=(1,-1,0,\ldots,0)$.% and $G^{D,DR\bigcdot WLS}_{\alpha}(O;\theta)=\{G^{DR\bigcdot WLS}_{0\alpha}(O;\mu_{0\alpha},\beta,\gamma),$ $G^{DR\bigcdot WLS}_{1\alpha}(O;\mu_{1\alpha},\beta,\gamma),$ $G^{reg\bigcdot WLS}_{\alpha}(O;\beta,\gamma), G^{ipw}(A,X;\gamma)\}^T$.
\end{proposition}

\subsection{Regression estimation with propensity based covariates}
In this section, a third DR estimator is considered which is constructed by including the inverse of the estimated propensity score in the regression model. Specifically, define the propensity based covariate DR estimator by 

\vspace{-0.8cm}
\begin{equation*}\label{eq: DR_picov_estimator}
  \widehat{Y}_i^{DR\bigcdot \pi\text{cov}}(a,\alpha)=N_i^{-1}\sum_{j=1}^{N_i}\Biggl\{\sum_{a_{i(-j)}}m_{ij}(a,a_{i(-j)},X_i;\hat\beta^{\pi\text{cov}})\pi(a_{i(-j)};\alpha)
 \Biggr\},
 \end{equation*}

\noindent where $\hat\beta^{\pi\text{cov}}=\hat\beta^{\pi\text{cov}}(\alpha)$ is obtained by solving $\sum_{i=1}^kG^{reg\bigcdot \pi\text{cov}}(A_i,X_i;\alpha,\beta,\hat\gamma)=0,$ 
%
%\vspace{-15mm}
%\begin{equation}\label{eq: ee_DR_picov}
%
%\end{equation}

\vspace{-6mm}
\[G^{reg\bigcdot \pi\text{cov}}(O_i;\beta,\gamma)=\tilde L_i^T\Lambda_i(A_i,X_i,1)\{Y_i- m_i(a,\tilde L_i;\beta,\gamma)\},\]
 $\tilde L_{ij}=\{1,A_{i(-j)},X_i,\pi(A_{i(-j)};\alpha)/f(a, A_{i(-j)}|X_i;\hat\gamma)\}$,  $\tilde L_i=(\tilde L_{i1}^T,\ldots,\tilde L_{iN_i}^T)^T$ and $\Lambda_i(A_i,X_i,1)=\text{diag}\bigl\{1(A_{i1}=a),\ldots,1(A_{iN_i}=a)\bigr\}$. That is, an additional covariate $\pi(A_{i(-j)};\alpha)/f(a, A_{i(-j)}|X_i;\hat\gamma)$ is included in the outcome regression model for $Y_{ij}$.
 
To gain some intuition for this type of DR estimator, note it is straightforward to show that conditional exchangeability implies $A_i\indep Y_{ij}(a_i)|f(a_i|X_i)$ and therefore
\begin{eqnarray*}
\mu_{a\alpha}&=& \sum_{a_{i(-j)}}E \{Y_{ij}(a,a_{i(-j)})\}\pi(a_{i(-j)};\alpha)\\
&=& \sum_{a_{i(-j)}}E[E \{Y_{ij}(a,a_{i(-j)})|f(a_i|X_i)\}]\pi(a_{i(-j)};\alpha)\\
%&=& \sum_{a_{i(-j)}}E[E \{Y_i|A_{ij}=a, A_{i(-j)}=a_{i(-j)},f(a,a_{i(-j)}|X_i)\}]\pi(a_{i(-j)};\alpha)\\
&=& \sum_{a_{i(-j)}}E[E \{Y_{ij}|A_{ij}=a, A_{i(-j)}=a_{i(-j)},f(a,a_{i(-j)}|X_i)\}]\pi(a_{i(-j)};\alpha).
\end{eqnarray*}

\noindent Hence, it is sufficient to model $E \{Y_{ij}|A_{ij}=a, A_{i(-j)}=a_{i(-j)},f(a,a_{i(-j)}|X_i)\}$. The DR property of $\widehat{Y}_i^{DR\bigcdot \pi\text{cov}}(a,\alpha)$ can be shown as in Section \ref{subsec: DR_WLS} by noting estimating equation \eqref{eq: ee_DR_WLS} is one of the estimating equations in $\sum_{i=1}^kG^{reg\bigcdot \pi\text{cov}}(O_i;\beta)=0 $ and thus the propensity based covariate DR estimator $\widehat{Y}_i^{DR\bigcdot \pi\text{cov}}(a,\alpha)$ can also be written in the same form as the bias correction DR estimator. %Additionally, when the outcome regression model is correct, the coefficient of the inverse weighted propensity score $\pi(a_{i(-j)};\alpha)/f(a, A_{i(-j)}|X_i;\hat\gamma)$ would be 0 since the inclusion of it as a covariate is just overfitting. 
This DR estimator can be viewed as a generalization of the DR estimator proposed by \citet{scharfstein1999adjusting} to the interference setting. 

To derive the asymptotic distribution, let $G^{DR\bigcdot \pi\text{cov}}_{a\alpha}(O;\mu_{0\alpha},\beta,\gamma)=\widehat Y_i^{DR\bigcdot \pi\text{cov}}(a,\alpha)-\mu$ and note $\hat \theta^{DR\bigcdot \pi\text{cov}}=\{\hat Y^{DR\bigcdot \pi\text{cov}}(0,\alpha),\hat Y^{DR\bigcdot \pi\text{cov}}(1,\alpha),\hat\beta,\hat\gamma\}$ is the solution to the estimating equation $$\sum_{i=1}^k G^{D,DR\bigcdot \pi\text{cov}}_{\alpha}(O_i;\theta)=0,$$ where $G^{ D,DR\bigcdot \pi\text{cov}}_{\alpha}(O;\theta) = \{G^{DR\bigcdot \pi\text{cov}}_{0\alpha}(O;\mu_{0\alpha},\beta,\gamma),G^{DR\bigcdot \pi\text{cov}}_{1\alpha}(O;\mu_{1\alpha},\beta,\gamma),G^{reg\bigcdot \pi\text{cov}}(O;\beta,\gamma),$\\ $G^{ipw}(A_i,X_i;\gamma)\}^T,$ $G^{DR\bigcdot \pi\text{cov}}_{a\alpha}(O_i;\mu,\beta,\gamma) =\widehat{Y}_i^{DR\bigcdot \pi\text{cov}}(a;\alpha,\beta,\gamma)- \mu$, and $\sum_i G^{ipw}(A_i,X_i;\gamma)=0$ is the estimating equation for the nuisance parameter estimate $\hat\gamma$. Propensity based covariate DR estimators can be constructed for the various causal effects, and these estimators are DR and asymptotically normal. This result for the direct effect estimator is stated formally by the following proposition.

\begin{proposition}\label{prop: DR_pi_cov}
If either $f(A_i|X_i;\gamma)$ or $m_{ij}(A_i,X_i;\beta)$ is correctly specified, then \\\emph{$k^{1/2}\{\widehat{\textsc{de}}^{{DR\bigcdot }\pi\text{cov}}(\alpha)-\overline{\textsc{de}}(\alpha)\}$} converges in distribution to $N(0,\Sigma_0^{D})$ as $k \to \infty$ where $
\Sigma^{D} = \tau U^{-1} V U^{-T}\tau^T,$ $U=-E\{\partial G_{\alpha}^{D,DR\bigcdot \pi\text{cov}}(O_i;\theta)/\partial \theta\}$, $V=E\{G_{\alpha}^{D,DR\bigcdot \pi\text{cov}}(O_i;\theta)^{\otimes 2}\}$, $\tau=(1,-1, 0,\ldots,0)$.% and $G^{D,DR\bigcdot \pi\text{cov}}_{\alpha}(O;\theta)=\{G^{reg}_{0\alpha}(O_i;\mu_{0\alpha},\beta,\gamma),$ $G^{reg}_{1\alpha}(O_i;\mu_{1\alpha},\beta,\gamma),$ $G^{DR\bigcdot \pi\text{cov}}(A_i,X_i;\beta,\gamma),$ $G^{ipw}(A_i,X_i;\gamma)\}^T$.
\end{proposition}

\section{Simulations}\label{sec: DR_simulation}
Simulations were conducted to assess the finite sample bias of the IPW, regression and DR estimators given in Sections \ref{sec: IPW_and_reg_estimator} and \ref{sec: DE_estimator} as well as to compare their efficiency and robustness when the models are either correct or mis-specified.
Simulations were conducted under four
scenarios: (i) both the propensity model and the outcome model were correct, (ii) the propensity model was wrong but the outcome model was correct, (iii) the propensity model was correct but the outcome model was wrong, and (iv) neither the propensity model or the outcome model was correct. For scenario (i), the simulation study was conducted in
the following steps:

\noindent \hangindent=1.5cm Step 1: We first generated a population with $m = 100$ groups and $N_i = 30$ individuals in each group. Covariates $X_{1ij}$ and $X_{2ij}$ were independently sampled from standard normal distribution and a Bernoulli distribution with probability 0.5, respectively.

\noindent \hangindent=1.5cm Step 2:  The treatment $A_{ij}$ was generated from the mixed effect logistic regression model $\mbox{logit}\{\Pr(A_{ij}=1|X_{1i}, X_{2i}, b_i) \} = 0.1 + 0.2 |X_{1ij}| + 0.2 |X_{1ij}| X_{2ij} + b_i$ where $b_{i}$ were independently and identically sampled from $N(0, 0.3)$.

\noindent \hangindent=1.5cm Step 3: The outcome $Y_{ij}$ was generated from $Y_{ij}=2 + 2A_{ij} + p(A_i) - 1.5 |X_{1ij}| + 2 X_{2ij} -3|X_{1ij}| X_{2ij}+\varepsilon_{ij}$ where $\varepsilon_{ij}$ independently and identically follow $N(0,1)$ and $p(A_i)$ was the proportion of treatment received among subjects in group $i$.

\noindent \hangindent=1.5cm Step 4: A correct outcome model $E\{Y_{ij}|X_{1i},X_{2i},A_i\}=\beta_0+
\beta_1A_{ij}+\beta_2p(A_i)+\beta_3|X_{1ij}|+\beta_4X_{2ij}+\beta_5 |X_{1ij}| X_{2ij}$ was fit and the outcome estimate $\hat{m}_{ij}(a_i,X_i)$ was calculated. 

\noindent \hangindent=1.5cm Step 5: A correct propensity model $\mbox{logit}\{\Pr(A_{ij}=1|X_{1i}, X_{2i}, b_i) \} = \gamma_0 + \gamma_1 |X_{1ij}| + \gamma_2 |X_{1ij}| X_{2ij} + b_i$ was fit to calculate the MLE $\hat{\gamma}$  and propensity score estimate ${f}(A_{ij}=1|X_i;\hat\gamma)$.

\noindent \hangindent=1.5cm Step 6: The IPW, regression and DR estimators were calculated according to Sections \ref{sec: IPW_and_reg_estimator} and \ref{sec: DE_estimator} with $\alpha_0 = 0.5$. 

%\noindent \hangindent=1.5cm Step 7: Step 1-6 were repeated for 

Scenario (ii) was carried out similar to Scenario (i) except Step 5 was replaced with

\noindent \hangindent=1.5cm Step 5$^{*}$:  A mis-specified propensity model $\mbox{logit}\{\Pr(A_{ij}=1|X_{1i}, X_{2i}, b_i) \} = \gamma_0 + \gamma_1 X_{1ij} +  b_i$ was fit to calculate the MLE $\hat{\gamma}$  and propensity score estimate ${f}(A_{ij}=1|X_i;\hat\gamma)$.

Scenarios (iii) was carried out similar to Scenario (i) except Step 4 was replaced with

\noindent \hangindent=1.5cm Step 4$^{*}$: A mis-specified outcome model $E\{Y_{ij}|X_{1i},X_{2i},A_i\}=\beta_0+
\beta_1A_{ij}+\beta_2p(A_i)+\beta_3X_{1ij}+\beta_4X_{2ij}$ was fit and the outcome regression estimate $\hat{m}_{ij}(a_i,X_i)$ was calculated.

\noindent  Scenario (iv) was carried out similar to Scenario (i) with Steps 4 and 5 replaced with Steps 4$^{*}$ and 5$^{*}$, respectively. The simulations were carried out 1400 times for the scenario with both component models correctly specified in order to accurately estimate confidence interval coverage to the second decimal. For the other scenarios where one or both of the component models was misspecified, the simulations were carried out 700 times. The propensity based covariate DR estimators involve integration of the propensity score over the random effect for each possible value of the treatment, and thus were excluded from the simulations due to computational burden under our specific models.

%The simulation result is presented in Figure \ref{fig:continuous}. As expected, when the $\pi$ model is correct, the IPW and the DR estimators have small bias while when the $\mu$ model is correct, the regression and DR estimators have small bias. Note that the DR estimator has a smaller variance than that of the IPW estimators when the $\pi$ model is correct; the DR estimator has a smaller variance than that of both the regression estimators when the $\mu$ model is correct. That is when one of the model is correct, the other model in DR estimator, although mis-specified, helps increase the efficiency. When both models are correct, the DR estimator has even smaller bias and variance. However, when both models are wrong, the DR estimator has as big or even bigger bias when using the IPW and regression estimators. This is also observed by Kang and Schafer (2007)\nocite{kang2007demystifying} who pointed out that `two wrong models are not better than one'.
%
%

\begin{figure}[h!]
\centering
\includegraphics[scale = .3]{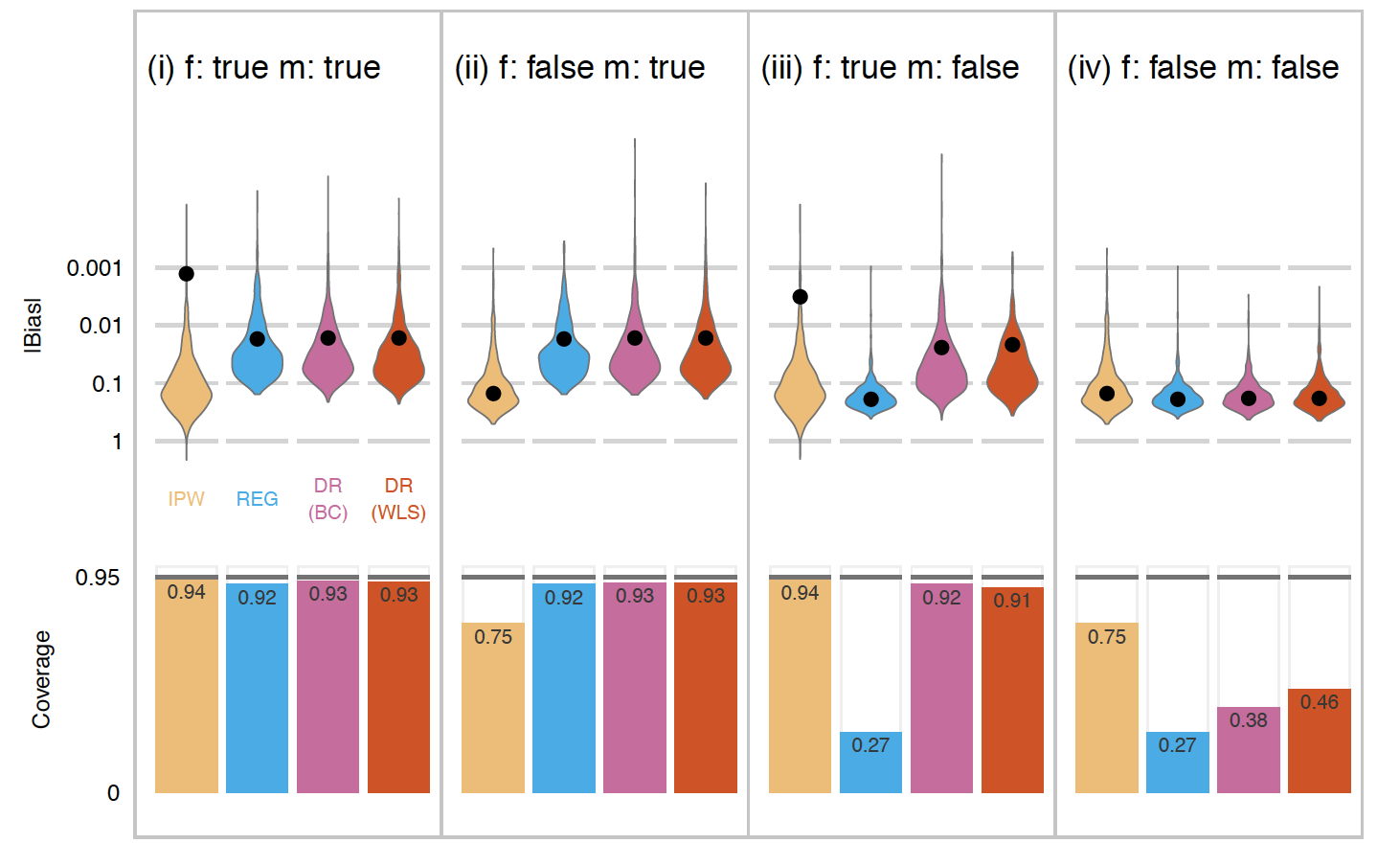}
\caption{Absolute bias and confidence interval coverage for the IPW, regression (REG), residual bias correction DR (DR (BC)) and weighted coefficients DR (DR (WLS)) estimators}
\label{fig:simulation_results}
\end{figure}

Simulation results are presented in Figure \ref{fig:simulation_results}. When the treatment model (i.e., propensity score) is correct, the IPW and the DR estimators have small bias while when the outcome model is correct, the regression and DR estimators have small bias. For example, for scenario (i), the bias of IPW, regression and residual bias correction DR and the weighted coefficients DR estimators are 0.001, -0.02, -0.02 and -0.02, respectively for $\mu_{1, 0.5}$. The residual bias correction DR and weighted coefficients DR estimators have smaller empirical variances (average standard error (ASE) = 0.053 and 0.055) than that of the IPW estimators (ASE = 0.21) when both the treatment and the outcome regression model model is correct. When the regression model is correctly specified, the regression estimator has the smallest variance. These comparison of variances align with the result without the interference as reported in Kang and Schafer (2007)\nocite{kang2007demystifying}. In our simulations, when both models are mis-specified, the DR estimators have substantial bias (-0.18 for both) as do the IPW and regression estimators (-0.15 and -0.18). %But this may just hold under this specific simulation setting. We do have observed in some scenarios, the DR estimators might have even bigger bias than using the IPW and regression estimators. This is also observed by Kang and Schafer (2007)\nocite{kang2007demystifying} who pointed out that two wrong models may not be better than one. 

Wald-type 95\% confidence intervals (CIs) were also constructed using the consistent variance estimators proposed in Section \ref{sec: DE_estimator}. Empirical coverages of the CIs are shown at the bottom of the Figure \ref{fig:simulation_results}. As expected, when the corresponding models are correctly specified  for the IPW and regression estimators, the corresponding CIs have coverage approximately equal to the 0.95. When either model is correctly specified for the DR estimators, the coverages are also approximately 0.95. When the models are mis-specified for the IPW and regression estimators or when neither of the models is correct for the DR estimators, the coverages are well below the nominal level. For example, when both models are wrong, the coverages are 0.75, 0.27, 0.38, and 0.46 for IPW, regression, residual bias correction DR and the weighted coefficients DR estimators, respectively.

\vspace{-0.4cm}
\section{Application}\label{sec: DR_dia}
A cholera vaccine trial was carried out in Matlab, Bangladesh in the late 1980s \citep{clemens1988field}. All children (2-15 yrs old) and women ($>$15 yrs old) were randomized with equal probability to one of three treatments: one of two types of cholera vaccine or a placebo. Following \citet{PerezHeydrich2014interference}, in the analysis presented here no distinction is made between the two cholera vaccines. Although the treatments were randomized, not all the eligible individuals participated. Those who did not participate in the randomized trial were followed for the primary outcome and included in the analysis; hence there is an observational aspect to these data. %the study contains an observational component.

Among the 121,982 eligible individuals, 49,300 individuals received at least two doses of vaccines. Previous analyses of these data suggest the risk of cholera among unvaccinated individuals was associated with the vaccine coverage in neighboring households or in their social network \citep{ali2005herd,root2011role}. \citet{PerezHeydrich2014interference} utilized the inverse probabilty weighted (IPW) estimators proposed by \citet{tchetgen2012causal} to assess the direct, indirect, total and overall causal effect of cholera vaccines. Here, we demonstrate the proposed DR estimators and compare to the IPW estimator and the outcome regression estimator. 

\citet{PerezHeydrich2014interference} used a spatial clustering algorithm to group individuals into 700 groups. Large groups cause significant computational burden for the outcome regression estimator. Since our primary purpose is a comparison of estimators, groups with more than 100 individuals were excluded from the analysis. This resulted in 14,589 individuals in 425 groups. Age in decades and distance to the nearest river were included as covariates in both the treatment and outcome models.

\begin{figure}[h!]
\centering
\includegraphics[scale = 1]{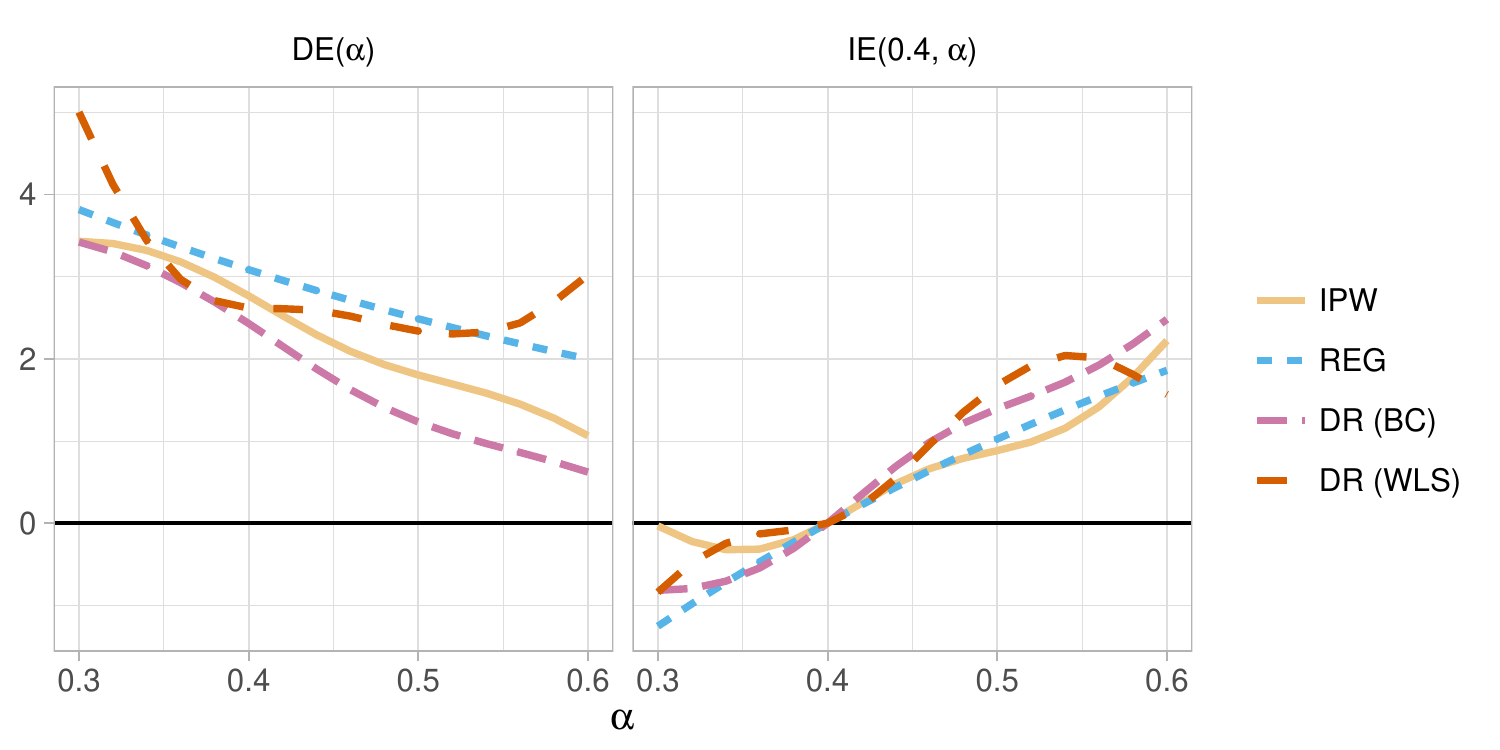}
\caption{Estimates of the direct ($\overline{\textsc{de}}(\alpha)$) and indirect ($\overline{\textsc{ie}}(0.4,\alpha)$) for the IPW, regression (REG), bias correction DR (DR (BC)) and weighted coefficient DR (DR (WLS)) estimators.}
\label{fig:cholera_results_DE_IE}
\end{figure}
%\clearpage
%\newpage

\begin{figure}[h!]
\centering
\includegraphics[scale = 1]{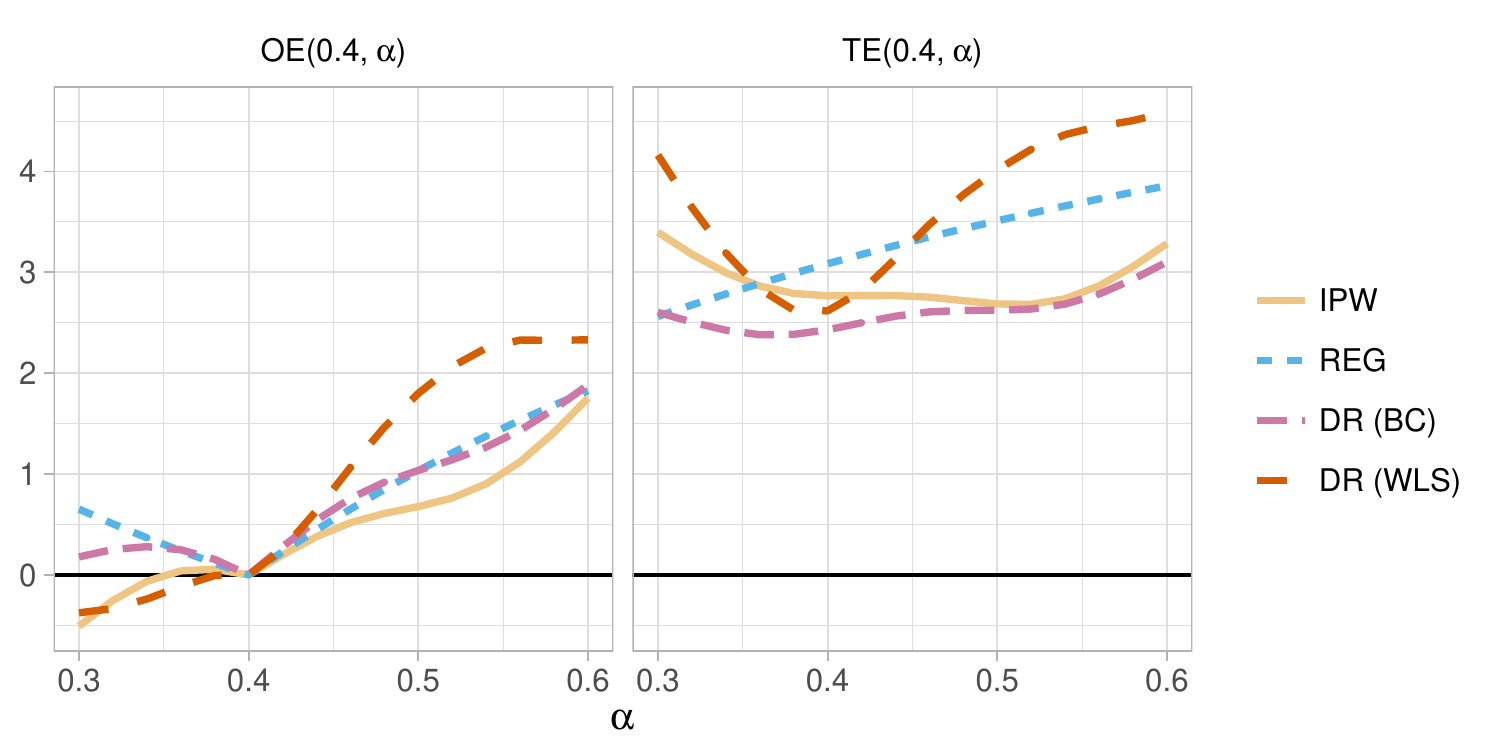}
\caption{Estimates of the total ($\overline{\textsc{te}}(0.4,\alpha)$) and overall effects ($\overline{\textsc{oe}}(0.4,\alpha)$) for the IPW, regression (REG), bias correction DR (DR (BC)) and weighted coefficient DR (DR (WLS)) estimators.}
\label{fig:cholera_results_TE_OE}
\end{figure}

Figures \ref{fig:cholera_results_DE_IE} and \ref{fig:cholera_results_TE_OE} compare the four estimators of the direct effect $\overline{\textsc{de}}(\alpha)$, indirect effect $\overline{\textsc{ie}}(0.4,\alpha)$, total effect $\overline{\textsc{te}}(0.4,\alpha)$ and overall effect $\overline{\textsc{oe}}(0.4,\alpha)$. The estimates and confidence intervals using the IPW, regression, and DR estimators are similar to \citet{PerezHeydrich2014interference}. While the point estimates using the weighted coefficients DR estimator are generally similar, the confidence intervals using this estimator are as much as 11 times wider than other methods. This could be due to the number of numerical approximations in estimating the covariance matrix. The estimating functions for this estimator correspond to 20 total target and nuisance parameters, compared to 8, 9, and 13 parameters for the IPW, regression, and residual bias DR estimators respectively.%whereas the IPW has 8, regression  9, and residual bias correction DR 13. %(\textbf{need to figure out why here})

As vaccine coverage $\alpha$ increases, the point estimate of the direct effect decreases among all estimators. For instance, when $\alpha=0.3$, the point estimate of the four estimators are approximately 3.4, 3.8, 3.4, and 5.0  for IPW, regression, biased correction DR and weighted coefficients DR estimators, respectively. This implies when the vaccine coverage is 30\%, we would expect to see about 3 or 4 fewer cases of cholera per 1000 person-years among the vaccinated individuals compared to unvaccinated ones. In comparison, when the vaccine coverage is around 60\%, the point estimates are approximately 1.1, 2.0, 0.6, and 3.0 for IPW, regression, biased correction DR and weighted coefficients DR estimators, respectively, indicating smaller change in the cases of cholera per 1000 person-years among the vaccinated individuals compared to unvaccinated ones. 

Unlike the direct effect, the indirect, total, and overall effect estimates incorporate interference, if present. The indirect effect estimate is negative when $\alpha<0.4$ and positive when $\alpha>0.4$, suggesting it is less likely for an unvaccinated individual to be infected when the vaccine coverage in their group is higher. The estimates of the total effect $\overline{\textsc{te}}(0.4,\alpha)$, which incorporate both the direct and indirect effects, are relatively constant as $\alpha$ increases, reflecting decreasing direct effect estimates offsetting increasing the indirect effect estimates.  The overall effect is in general higher for higher coverage groups as compared with lower coverage groups. For example,  when the vaccine coverage is 30\% compared to 40\%, the overall effect is estimated to be -0.51, 0.65, 0.18, and -0.38 for IPW, regression, biased correction DR and weighted coefficients DR estimators, respectively, while when the vaccine coverage is 60\%, those are 1.6, 1.8, 1.9 and 2.3. Point estimates and 95\% Wald-type confidence intervals of the IPW, regression and DR estimators for various effects are given in Table \ref{tb: real_data_CI}.

\begin{table}
\caption{Estimates and 95\% Wald-type confidence intervals of the direct ($\overline{\textsc{de}}(\alpha)$), indirect ($\overline{\textsc{ie}}(0.4,\alpha)$), total ($\overline{\textsc{te}}(0.4,\alpha)$) and overall effects ($\overline{\textsc{oe}}(0.4,\alpha)$) for the IPW, regression (REG), bias correction DR (DR (BC)) and weighted coefficient DR (DR (WLS)) estimators\label{tb: real_data_CI}}
\begin{tabular}{lllll}
\toprule
 & IPW & REG & DR (BC) & DR (WLS)\\
%\midrule
$\overline{\textsc{de}}(\alpha)$ &&&&\\
 $\alpha=0.30$ & 3.4 (0.0, 6.9) & 3.8 (1.0, 6.7) & 3.4 ($-$0.2, 7.0) & 5.0 ($-$1.2, 11.2)\\
 $\alpha=0.44$ & 2.3 ($-$1.1, 5.7) & 2.8 (0.8, 4.8) & 1.9 ($-$1.5, 5.3) & 2.6 ($-$2.1, 7.3)\\
 $\alpha=0.60$ & 1.1 ($-$1.8, 3.9) & 2.0 (0.5, 3.5) & 0.6 ($-$2.3, 3.5) & 3.0 ($-$1.3, 7.3)\\
%\hline
$\overline{\textsc{ie}}(0.4,\alpha)$ &&&&\\
 $\alpha=0.30$ & 0.0 ($-$1.9, 1.8) & $-$1.3 ($-$2.4, $-$0.1) & $-$0.8 ($-$2.5, 0.9) & $-$0.8 ($-$6.0, 4.3)\\
 $\alpha=0.44$ & 0.5 ($-$0.2, 1.2) & 0.4 (0.1, 0.8) & 0.7 (0.1, 1.3) & 0.5 ($-$2.7, 3.8)\\
$\alpha=0.60$ & 2.2 ($-$0.2, 4.6) & 1.9 (0.5, 3.2) & 2.5 (0.2, 4.8) & 1.6 ($-$5.8, 0.0)\\
%\hline
$\overline{\textsc{te}}(0.4,\alpha)$&&&&\\
 $\alpha=0.30$ & 3.4 (0.0, 6.8) & 2.6 (0.1, 5.0) & 2.6 ($-$1.0, 6.2) & 4.2 ($-$2.4, 10.7)\\
 $\alpha= 0.44$ & 2.8 ($-$0.7, 6.3) & 3.3 (1.1, 5.4) & 2.6 ($-$0.9, 6.1) & 3.1 ($-$3.4, 9.7)\\
 $\alpha=0.60$ & 3.3 ($-$0.1, 6.6) & 3.9 (1.8, 5.9) & 3.1 ($-$0.2, 6.4) & 4.6 ($-$1.7, 10.9)\\
%\hline
$\overline{\textsc{oe}}(0.4,\alpha)$&&&&\\
 $\alpha=0.30$ & $-$0.5 ($-$1.8, 0.8) & 0.7 ($-$0.2, 1.5) & 0.2 ($-$1.0, 1.4) & $-$0.4 ($-$3.9, 3.1)\\
 $\alpha=0.44$ & 0.4 ($-$0.14, 0.9) & 0.4 (0.2, 0.7) & 0.5 (0.1, 1.0) & 0.6 ($-$1.4, 2.7)\\
 $\alpha=0.60$ & 1.8 (0.1, 3.4) & 1.8 (0.8, 2.8) & 1.9 (0.3, 3.5) & 2.3 ($-$2.0, 6.7)\\
\bottomrule
\end{tabular}
\end{table}

%\textbf{I excluded the confidence intervals for now. They're difficult to visualize when overlaying the  4 methods. OK. Shall we just report the point estimate in figure? It is indeed hard to see for all the estimators. Maybe we could mention the CI in the text.}
\vspace{-.3cm}
\section{Discussion}\label{sec: DR_discussion}
\vspace{3mm}

In this paper several DR estimators are proposed for causal effects in the presence of partial interference. The estimators are shown to be consistent and asymptotically normal if either the propensity model or the outcome regression model, but not necessarily both, is correctly specified. Empirical results demonstrate the DR property of the proposed estimators and possible efficiency gains over a previously proposed IPW estimator when both models are correctly specified.

Application of the proposed methods to the cholera vaccine study provides robust evidence corroborating previous analyses that population-level vaccination affords a protective indirect effect to unvaccinated individuals. As in \citet{PerezHeydrich2014interference}, the analysis presented here demonstrates how considering only the direct effect of a vaccine may fail to capture the totality of effects afforded by vaccination at the population level. Note the formulation in this paper considers the direct effect to be a function of vaccine coverage. That is, there is not a single direct effect, but rather a direct effect curve which describes the individual effect of vaccination for a given level of vaccine coverage in the population. Traditionally the direct effect of a vaccine refers to the direct protection for a vaccinated individual owing only to vaccine-induced immunity in that individual \citep{clemens2011new}; in the current formulation, this would correspond to the direct effect when the level of vaccine coverage is 0\%. In settings where interference is present, the direct effect curve may vary with vaccine coverage, in which case simple analyses about the direct effect from studies with high levels of vaccine coverage may mislead about the standard interpretation of the direct effect of a vaccine. On the other hand, the methods developed in this paper permit robust inference of the direct effect curve, providing public health officials and policy makers a more complete picture of how the individual effect of vaccination changes with vaccine coverage.  

There are several areas of possible future research related to the methods developed here. For instance, whether any of the DR estimators proposed are semiparametric efficient remains to be investigated. Future research could also entail extensions of the proposed DR estimators to the setting where there is general interference, similar to the IPW estimator for general interference proposed by \citet{liu2016inverse}. In the absence of interference, DR estimator have the appealing property of achieving parametric rates of convergence (i.e., $n^{-1/2}$)  even if the working outcome and propensity models are non-parametric provided the estimators of the working model parameters (i.e., nuisance parameters) converge at rate greater than $n^{-1/4}$ \citep{naimi2017nonparametric}, allowing data-adaptive methods for fitting the working models. Whether the DR estimators proposed in this paper also have this property remains to be determined.

\section*{Acknowledgments}

This work was supported by NIH grant R01 AI085073.

%{\it Conflict of Interest}: None declared.

\bibliographystyle{apsr}
%\bibliography{REF FILE}
\bibliography{mybib}

\vspace{10mm}
\noindent \textbf{\Large Appendix}
\section*{}
\vspace{-20mm}
\subsection*{Proof of Proposition \ref{thm: DR_CAN}}
%Note the true parameter value $\theta=(\mu_{0\alpha},\mu_{1\alpha},\beta,\gamma)$ is the solution to the vector equation $\int G^{D, DR\bigcdot BC}_{\alpha}(O;\theta) dF(o) = 0$.

 To prove the doubly robustness of bias correction DR estimators, 
% assume %under certain regularity conditions, 
% there exists $\gamma^{*}$ such that $\hat{\gamma}\xrightarrow{p}\gamma^{*}$ as $k\rightarrow\infty$, that is, the estimator $\hat{\gamma}$ will converge in probability to some constant regardless of whether the propensity model is correct or not. Similarly, assume there exist $\beta^{*}$ such that $\hat{\beta}\xrightarrow{p}\beta^{*}$ as $k\rightarrow\infty$. L
 let $\gamma_0$ and $\beta_0$ denote the true values of the parameters in the propensity score and outcome regression models. Define $\beta^*$ to be such that $E\{G^{DR\bigcdot BC}_{\beta}(O_i;\beta^*)\}=0$; note here and below the expectation is taken with respect to the true parameters. 
Likewise, define $\gamma^*$ to be such that $E\{{G^{DR\bigcdot BC}_{\gamma}(O_i;\gamma^*)}\}=0$.
  If the propensity score (or outcome regression model) is correctly specified, then $\gamma^{*}=\gamma_0$ (or $\beta^{*}=\beta_0$).

%$G^{D,DR\bigcdot BC}_{\alpha}(y,a,x;\theta)=(G^{DR\bigcdot BC}_{0\alpha}(O_i;\mu_{0\alpha},\beta,\gamma),G^{DR\bigcdot BC}_{1\alpha}(O_i;\mu_{1\alpha},\beta,\gamma),G^{DR\bigcdot BC}(O_i;\beta,\gamma))^T$  and thus $E\{G^{D,DR\bigcdot BC}_{\alpha}(O_i;\theta^{*})\}=0$ G^{DR\bigcdot BC}_{a\alpha}(O_i;\mu,\beta,\gamma) =\widehat{Y}_i^{DR\bigcdot BC}(a;\alpha)- \mu

%\vspace{-.2cm}
If $\gamma^{*}=\gamma_0$, then ${f}(A_i|X_i;\gamma^{*})=f(A_i|X_i;\gamma_0)$ and thus

\vspace{-0.4cm}
\begin{equation*}
  E\Biggl\{\frac{1(A_{ij}=a)Y_{ij}(A_i)}{{f}(A_i|X_i;\gamma^{*})}\pi(A_{i(-j)};\alpha)\Biggr\}=m_{ij}(a,\alpha).
\end{equation*}

\vspace{-0.3cm}
\noindent Also note that

\vspace{-1cm}
\begin{eqnarray*}
&&  E\Biggl\{\sum_{a_{i(-j)}}m_{ij}(a_i,X_i;\beta^{*})\pi(a_{i(-j)};\alpha)
  -
  \frac{1(A_{ij}=a)m_{ij}(A_i,X_i;\beta^{*})}{f(A_i|X_i;\gamma^{*})}\pi(A_{i(-j)};\alpha)\Biggr\}\\
&=&  E\Biggl\{\sum_{a_{i(-j)}}m_{ij}(a_i,X_i;\beta^{\ast})\pi(a_{i(-j)};\alpha)\Biggr\}\\
&&  - E\Biggl\{\sum_{a_{i(-j)}}\frac{m_{ij}(a,a_{i(-j)},X_i;\beta^{\ast})}{{f}(a,a_{i(-j)}|X_i;\gamma_0)}
  \pi(a_{i(-j)};\alpha)\Pr(a,a_{i(-j)}|X_i)\Biggr\}\\
&=&  E\Biggl\{\sum_{a_{i(-j)}}m_{ij}(a_i,X_i;\beta^{\ast})\pi(a_{i(-j)};\alpha)\Biggr\}
  - E\Biggl\{\sum_{a_{i(-j)}}m_{ij}(a,a_{i(-j)},X_i;\beta^{\ast})
  \pi(a_{i(-j)};\alpha)\Biggr\}
=0,
  \end{eqnarray*}

\noindent which implies $E\{G^{DR\bigcdot BC}_{a\alpha}(O_i;\mu_{a\alpha},\beta^*,\gamma^*)\}=0$.

%\vspace{-.3cm}
If $\beta^{*}=\beta_0$, then $m_{ij}(a_i,X_i;\beta^{*})=m_{ij}(a_i,X_i;\beta_0)$  and thus

\vspace{-.4cm}
\begin{equation*}
 E\Biggl\{\sum_{a_{i(-j)}}m_{ij}(a,a_{i(-j)},X_i;\beta^{*})\pi(a_{i(-j)};\alpha)\Biggr\}=m_{ij}(a,\alpha).
\end{equation*}

\vspace{-.4cm}
\noindent Also note that

\vspace{-1cm}
\begin{eqnarray*}
  E\Biggl\{\frac{1(A_{ij}=a)\{Y_{ij}(A_i)-m_{ij}(A_i,X_i;\beta^*)\}}{{f}(A_i|X_i;\gamma^{*})}\pi(A_{i(-j)};\alpha)\Biggr\}
=0,
\end{eqnarray*}

\noindent which implies $E\{G^{DR\bigcdot BC}_{a\alpha}(O_i;\mu_{a\alpha},\beta^{*},\gamma^{*})\}=0$. Thus, $E\{G^{DR\bigcdot BC}_{a\alpha}(O_i;\mu_{a\alpha},\beta^{*},\gamma^{*})\}=0$ when either propensity score or outcome regression is correctly specified. %Assuming $\sum_{v=1}^k G^{D,DR\bigcdot BC}_{\alpha}(O_i;\theta)=0$ has a unique solution, it follows that $\widehat{Y}_i^{DR\bigcdot BC}(a;\alpha)\xrightarrow{p}\mu_{a\alpha}$ and thus $\widehat{\textsc{de}}^{DR\bigcdot BC}(\alpha)\xrightarrow{p}\overline{\textsc{de}}(\alpha)$ when either the propensity score model or the outcome regression model is correctly specified. Therefore, $\widehat{\textsc{de}}^{DR\bigcdot BC}(\alpha)$ is doubly robust.

%To derive the asymptotic normality, note the true parameter value $\theta=(\mu_{0\alpha},\mu_{1\alpha},\beta_0,\gamma_0)$ is the solution to the vector equation $\int G^{D, DR\bigcdot BC}_{\alpha}(O;\theta) dF(o) = 0$. Assume $E(Y)<\infty$, $E(Y^2)<\infty$, and $E(N_i)<\infty$, implying $E||G^{D,DR\bigcdot BC}_{\alpha}(O_i;\theta)||^2<\infty$. Additionally, assume $E\{\partial G^{D,DR\bigcdot BC}_{\alpha}(O_i;\theta)/\partial \theta\}$ exists and is non-singular, and that the second order derivatives of $G^{D,DR\bigcdot BC}_{\alpha}(O_i;\theta)$ are dominated by a fixed integrable function for every $\theta$ in a neighborhood of true value $\theta_0$. By Theorem 5.4.1 of \cite{van2000asymptotic}, 

Let $\theta_0=(\mu_{0\alpha}, \mu_{1\alpha},\beta^*,\gamma^*)$. Then by standard estimating equation theory (Stefanski and Boos, 2002; van der Vaart, A. 1998 Ch.\ 5) \nocite{stefanski2002calculus,van2000asymptotic} it follows that under suitable regularity conditions that 
\emph{$k^{1/2}\{\hat{\theta}^{DR\bigcdot BC}-\theta_0\}$} converges in distribution to $N(0,\Sigma)$ as $k \to \infty$ where $\Sigma = U^{-1} V U^{-T},$ $U=-E\{\partial G_{\alpha}^{D,DR\cdot BC}(O_i;\theta)/\partial \theta\}$ and $V=E\{G_{\alpha}^{D,DR\bigcdot BC}(O_i;\theta)^{\otimes 2}\}$. Asymptotic normality of $\widehat{\textsc{de}}^{DR\bigcdot BC}(\alpha)$ follows from the delta method.

\subsection*{Proof of Proposition \ref{prop: DR_WLS}}

%\begin{proof}
As in the proof of Proposition \ref{thm: DR_CAN}, let $\gamma_0$ and $\beta_0$ denote the true values of the parameters in the propensity score and outcome regression models. Define $\beta^*$  and $ \gamma^*$ to be such that $E\{G^{reg\bigcdot WLS}_{\alpha}(O_i;\beta^*,\gamma^*)\}=0$ and $E\{G^{ipw}(A_i,X_i;\gamma^*)\}=0$.  If $\gamma^{*}=\gamma_0$, the $j^{th}$ element of $E\{G^{DR\bigcdot WLS}(O_i;\alpha,\beta^{*},\gamma^{*})\}$ equals

\begin{eqnarray*}
%&&E\Biggl\{\frac{1(A_{ij}=a)\{Y_{ij}(A_i)-m_{ij}(A_i,X_i;\hat\beta^{WLS})\}}{{f}(A_i|X_i;\hat\gamma)}
%  \pi(A_{i(-j)};\alpha)\biggr\}\\
%  &\xrightarrow{p}&
  &&E\Biggl\{\frac{1(A_{ij}=a)\{Y_{ij}(A_i)-m_{ij}(A_i,X_i;\beta^{\ast})\}}{{f}(A_i|X_i;\gamma^{*})}
  \pi(A_{i(-j)};\alpha)\biggr\}\\
    &=&E\Biggl\{\sum_{a_{i(-j)}}\frac{\{Y_{ij}(a,a_{i(-j)})-m_{ij}(a,a_{i(-j)},X_i;\beta^{\ast})\}}{{f}(a_i|X_i;\gamma_0)}
  \pi(a_{i(-j)};\alpha){f}(a_i|X_i;\gamma_0)\biggr\}\\
      &=&E\Biggl\{\sum_{a_{i(-j)}}\{Y_{ij}(a,a_{i(-j)})-m_{ij}(a,a_{i(-j)},X_i;\beta^{\ast})\}
  \pi(a_{i(-j)};\alpha)\biggr\}\\
  &=&\mu_{a\alpha}-E\Biggl\{\sum_{a_{i(-j)}}m_{ij}(a,a_{i(-j)},X_i;\beta^{\ast})
  \pi(a_{i(-j)};\alpha)\biggr\}.
\end{eqnarray*}

\noindent Thus, 

\begin{eqnarray*}
 E\{G^{DR\bigcdot WLS}_{a\alpha}(O_i;\mu_{a\alpha},\beta^*,\gamma^*)\}%&=&N_i^{-1}\sum_{j=1}^{N_i}\Biggl\{\sum_{a_{i(-j)}}m_{ij}(a,a_{i(-j)},X_i;\hat\beta^{WLS})\pi(a_{i(-j)};\alpha)
% \Biggr\}-\mu_{a\alpha}\\
 &=&E\Biggl\{\sum_{a_{i(-j)}}m_{ij}(a,a_{i(-j)},X_i;\beta^{\ast})
  \pi(a_{i(-j)};\alpha)\biggr\}-\mu_{a\alpha}=0.
\end{eqnarray*}

%\noindent Thus, $f(A_i|X_i;\gamma)$ is correctly specified, $\widehat{\textsc{de}}^{{DR\bigcdot WLS}}(\alpha)\xrightarrow{p}\overline{\textsc{de}}(\alpha)$.

If $\beta^{*}=\beta_0$, then $m_{ij}(a_i,X_i;\beta^{*})=m_{ij}(a_i,X_i;\beta_0)$  and thus, $m_{ij}(a,a_{i(-j)},X_i;\beta_0)=E\{Y_{ij}(a,a_{i(-j)})|X_i\}$. Note

\begin{eqnarray*}
%&&E\Biggl\{\frac{1(A_{ij}=a)\{Y_{ij}(A_i)-m_{ij}(A_i,X_i;\hat\beta^{WLS})\}}{{f}(A_i|X_i;\hat\gamma)}
  %\pi(A_{i(-j)};\alpha)\biggr\}\\
  %&\xrightarrow{p}&
  &&E\Biggl\{\frac{1(A_{ij}=a)\{Y_{ij}(A_i)-m_{ij}(A_i,X_i;\beta_0)\}}{{f}(A_i|X_i;\gamma^{\ast})}
  \pi(A_{i(-j)};\alpha)\biggr\}\\
    &=&E\Biggl\{\sum_{a_{i(-j)}}\frac{\{Y_{ij}(a,a_{i(-j)})-m_{ij}(a,a_{i(-j)},X_i;\beta_0)\}}{{f}(a_i|X_i;\gamma^{\ast})}
  \pi(a_{i(-j)};\alpha){f}(a_i|X_i;\gamma_0)\biggr\}\\
    &=&\sum_{a_{i(-j)}}E\Biggl\{\frac{\bigl\{E\{Y_{ij}(a,a_{i(-j)})|X_i\}-m_{ij}(a,a_{i(-j)},X_i;\beta_0)\bigr\}}{{f}(a_i|X_i;\gamma^{\ast})}
  {f}(a_i|X_i;\gamma_0)\biggr\}\pi(a_{i(-j)};\alpha)\\
  &=&0.
\end{eqnarray*}

\noindent And therefore,

\begin{eqnarray*}
 E\{G^{DR\bigcdot WLS}_{a\alpha}(O_i;\mu_{a\alpha},\beta^*,\gamma^*)\}%&=&N_i^{-1}\sum_{j=1}^{N_i}\Biggl\{\sum_{a_{i(-j)}}m_{ij}(a,a_{i(-j)},X_i;\hat\beta^{WLS})\pi(a_{i(-j)};\alpha)
% \Biggr\}-\mu_{a\alpha}\\
 &=&E\Biggl\{\sum_{a_{i(-j)}}m_{ij}(a,a_{i(-j)},X_i;\beta^{\ast})
  \pi(a_{i(-j)};\alpha)\biggr\}-\mu_{a\alpha}=0.
\end{eqnarray*}

\noindent Thus, when  either propensity score or outcome regression model is correctly specified, \\$ E\{G^{DR\bigcdot WLS}_{a\alpha}(O_i;\mu_{a\alpha},\beta^*,\gamma^*)\}=0$. Asymptotic normality of $\widehat{\textsc{de}}^{DR\bigcdot WLS}(\alpha)$ follows along the same lines as the proof of Proposition \ref{thm: DR_CAN}.
 
%To derive the asymptotic normality, assume $E(Y)<\infty$, $E(Y^2)<\infty$, and $E(N_i)<\infty$, implying $E||G^{D,DR\bigcdot WLS}_{\alpha}(O_i;\theta)||^2<\infty$. Additionally, $E\{\partial G^{D,DR\bigcdot WLS}_{\alpha}(O_i;\theta^{*})/\partial \theta^{*}\}$ exists and is non-singular. Assume that the second order derivatives of $G^{D,DR\bigcdot WLS}_{\alpha}(O_i;\theta^{*})$ are dominated by a fixed integrable function for every $\theta^{*}$ in a neighborhood of $\theta_0$. By Theorem 5.4.1 of \cite{van2000asymptotic}, \emph{$k^{1/2}\{\hat{\theta}^{DR\bigcdot WLS}-\theta_0\}$} converges in distribution to $N(0,\Sigma)$ as $k \to \infty$ where $\Sigma = U^{-1} V U^{-T},$ $U=-E\{\partial G_{\alpha}^{D,DR\cdot WLS}(O_i;\theta)/\partial \theta\}$ and $V=E\{G_{\alpha}^{D,DR\bigcdot WLS}(O_i;\theta)^{\otimes 2}\}$. The asymptotic normality of $\widehat{\textsc{de}}^{DR\bigcdot WLS}(\alpha)$ follows from delta method.

%\end{proof}DR\bigcdot }\pi\text{cov}
\subsection*{Proof of Proposition \ref{prop: DR_pi_cov}}

The proof of the DR property of $\widehat{\textsc{de}}^{{DR\bigcdot \pi\text{cov}}}(\alpha)$ is similar to that of $\widehat{\textsc{de}}^{{DR\bigcdot WLS}}(\alpha)$ in Proposition \ref{prop: DR_WLS} since estimating equation \eqref{eq: ee_DR_WLS} is also contained in $\sum_{i=1}^kG^{reg\bigcdot \pi\text{cov}}(O_i;\beta)=0$. Specifically, if $\gamma^{*}=\gamma_0$, following a same argument as in the proof of Proposition \ref{prop: DR_WLS}, we have $E\{G^{DR\bigcdot \pi\text{cov}}_{a\alpha}(O_i;\mu,\beta^*,\gamma^*)\}=0$. If $\beta^{*}=\beta_0$ then the coefficient for $\pi(A_{i(-j)};\alpha)/f(a, A_{i(-j)}|X_i;\gamma^*)$ is 0, hence, we have $m_{ij}(a,\tilde L_i;\beta^{*},\gamma^{*})=m_{ij}(a_i,X_i;\beta_0)$. Following a similar argument, we have when $m_{ij}(A_i,X_i;\beta)$ is correctly specified, \\$E\{G^{DR\bigcdot \pi\text{cov}}_{a\alpha}(O_i;\mu,\beta^*,\gamma^*)\}=0$. Thus, $\widehat{\textsc{de}}^{{DR\bigcdot \pi\text{cov}}}(\alpha)$ is doubly robust. Asymptotic normality of $\widehat{\textsc{de}}^{DR\bigcdot \pi\text{cov}}(\alpha)$ follows as in the proof  of Proposition \ref{thm: DR_CAN}.

%The asymptotic normality of $\widehat{\textsc{de}}^{DR\bigcdot \pi\text{cov}}(\alpha)$ could be derived similarly as for $\widehat{\textsc{de}}^{DR\bigcdot BC}(\alpha)$ and $\widehat{\textsc{de}}^{DR\bigcdot WLS}(\alpha)$. Specifically, we assume $E(Y)<\infty$, $E(Y^2)<\infty$, and that $E(N_i)<\infty$, then we have $E||G^{D,DR\bigcdot \pi\text{cov}}_{\alpha}(O_i;\theta)||^2<\infty$. Additionally, assume that the second order derivatives of $G^{D,DR\bigcdot \pi\text{cov}}_{\alpha}(O_i;\theta^{*})$ are dominated by a fixed integrable function for every $\theta^{*}$ in a neighborhood of $\theta_0$. By Theorem 5.4.1 of \cite{van2000asymptotic}, we have \emph{$k^{1/2}\{\hat{\theta}^{DR\bigcdot \pi\text{cov}}-\theta_0\}$} converges in distribution to $N(0,\Sigma)$ as $k \to \infty$ where $\Sigma = U^{-1} V U^{-T},$ $U=-E\{\partial G_{\alpha}^{D,DR\bigcdot \pi\text{cov}}(O_i;\theta)/\partial \theta\}$ and $V=E\{G_{\alpha}^{D,DR\bigcdot \pi\text{cov}}(O_i;\theta)^{\otimes 2}\}$. The asymptotic normality of $\widehat{\textsc{de}}^{DR\bigcdot \pi\text{cov}}(\alpha)$ follows from delta method.
%$E\{\partial G^{D,DR\bigcdot \pi\text{cov}}_{\alpha}(O_i;\theta^{*})/\partial \theta^{*}\}$ exists and is non-singular. At

%\end{appendix}

\end{document}